# An Empirical Study and Analysis of the Dynamic Load Balancing Techniques Used in Parallel Computing Systems

Ardhendu Mandal[§] and Subhas Chandra Pal[†]

Department of Computer Science and Application, University of North Bengal, Raja Rammohanpur,

P.O. North Bengal University, Dist. Darjeeling, West Bengal, Pin-734013, India.

am.csa.nbu@gmail.com[§], schpal@rediffmail.com[†]

*Abstract-* A parallel computer system is a collection of processing elements that communicate and cooperate to solve large computational problems efficiently. To achieve this, at first the large computational problem is partitioned into several tasks with different work-loads and then are assigned to the different processing elements for computation. Distribution of the work load is known as Load Balancing. An appropriate distribution of work-loads across the various processing elements is very important as disproportional workloads can eliminate the performance benefit of parallelizing the job. Hence, load balancing on parallel systems is a critical and challenging activity. Load balancing algorithms can be broadly categorized as static or dynamic. Static load balancing algorithms distribute the tasks to processing elements at compile time, while dynamic algorithms bind tasks to processing elements at run time. This paper explains only the different dynamic load balancing techniques in brief used in parallel systems and concluding with the comparative performance analysis result of these algorithms.

**Keywords:** Parallel Computer, Parallel Programming, Partitioning, Load Balancing, Static Load Balancing, Dynamic Load Balancing, load balancer.

## I. Introduction

Most large scientific computations are now carried out on parallel computers. A *parallel computer* is a computing system with multiple number of processing elements that communicate and cooperate to solve large problems efficiently. Achieving the improved performance objective in parallel systems with general sequential programming is insufficient.

With multiple processing elements, the larger job as to be divided in smaller tasks to be distributed over these processing elements to carry out the partial results in parallel. This originates the concepts of *parallel programming*. Hence, creating parallel programs involves decomposition i.e. *partitioning* the overall computation into several tasks and then *assigning* these smaller tasks to multiple processing elements. The number of tasks generated by the partitioning may not be equal to the processors, thus a processor may be idle or loaded with multiple processes. In addition, although the number of tasks and number of processing elements is equal, often it doesn't ensure the optimized performance as work-load to individual processors may be unequal. A further challenge remains alive with equal work-load per tasks when computational power of individual processing elements varies. The module responsible for performing the load balancing job is called the *load balancer*. The technique of managing work load on processing elements is known as *load balancing*.

The rest of the paper is organized by starting with different load balancing techniques with detail specific discussion of dynamic load balancing techniques only, followed by different dynamic load balancing techniques available and concluded with their study of comparative performance results. Load balancing algorithms aims to equalize the workload among nodes.

## II. Goals of Load Balancing

The load balancing of an application has a direct impact on the speedup to be achieved as well as in the performance of the parallel system [1, 2].

Redistribution of balanced work-load by means of tasks and minimizing the inter process communication needs with optimal resource utilization and job response time are the primary optimization objective of load balancing. Hence, improving the performance of parallel computers by equalizing the workloads of processing elements is the *aim of load balancing*.

Some of the main goals of a load balancing algorithm, as pointed out by [3] are to:

(1) **Performance Improvement:** Achieve a greater overall improvement in system performance at a reasonable cost, e.g., reduce task response time while keeping acceptable delays;

(2) **Job Equality:** To treat all jobs in the system equally regardless of their origin;

(3) **Fault Tolerance:** To have performance endurance under partial failure in the system;

(4) **Modifiability:** Have the ability to modify itself in accordance with any changes or expand in the distributed system configuration; and

(5) **System stability:** The ability to account for emergency situations such as sudden surge of arrivals so that system performance does not deteriorate beyond a certain threshold while preventing nodes of the distributed system from spending too much time passing up jobs among themselves instead of executing these jobs.

## III. Types of Load Balancing Strategies

Various strategies and algorithms have been proposed, implemented and classified in a number of studies [4, 5, 6]. Broadly, load balancing is a kind of scheduling optimization problem.

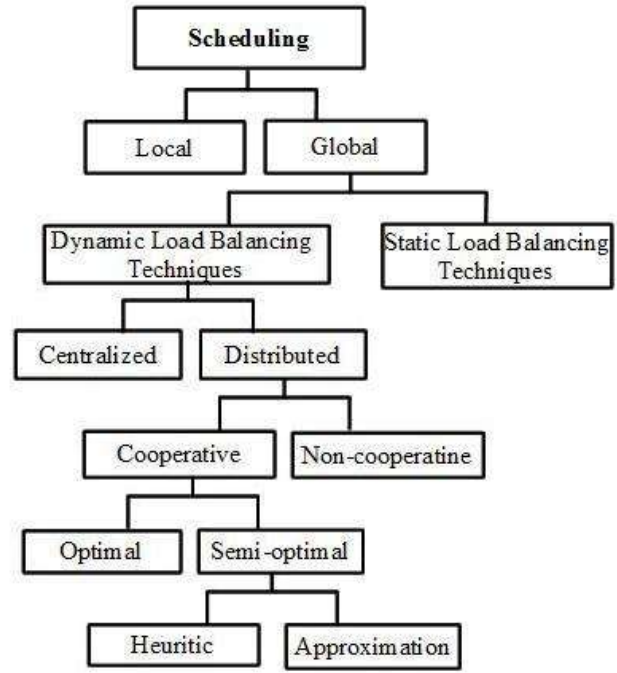

Figure 1. Different Types of Load Balancing Techniques.

The load balancing strategy may be determined by *inspection*, such as with a rectangular lattice of grid points split into smaller rectangles, so that the load balancing problem is solved before the program is written. Depending on the *information used* in load balancing decision, it can be divided into two broad categories i.e. g*lobal* or *local* policies [6]. In *global* policies, the load balancer uses the performance profiles of all available workstations. In *local* policies workstations are partitioned into different groups. The benefit in a local scheme is that performance profile information is only exchanged within the local group. The choice of a global or local policy depends on the behavior an application will exhibit. Depending on the time to bind the tasks to processing elements, load balancing algorithms can be further categorized as static or dynamic [7]. The non-trivial static load balancing algorithms distribute the tasks to processing elements at compile time, while dynamic algorithms bind tasks to processing elements at run time. Static load balancing algorithms rely on the estimate execution times of the tasks and inter-process communication requirement. It is not satisfactory for parallel programs that are of the dynamic and/or unpredictable kind. Consequently in dynamic load balancing, tasks are generated and destroyed without a pattern at run time. Further, depending on the location where the load balancing decision is carried out i.e. the resident of the load balancer, these can be further classified either as centralized or distributed load balancing. The case when the load balancer resides at the master node is called centralized load balancing policy, otherwise if the same resides at all the workstations under consideration is called the distributed load balancing policy. Further, in quasi-dynamic, the circumstances determining the optimal balance change during program execution, but discretely and infrequently. Because the change is discrete, the load balance problem and hence its solution remain the same until the next change. If these changes are infrequent enough, any savings made in the subsequent computation make up for the time spent solving the load balancing problem. The difference between this and the static case is that the load balancing must be carried out in parallel to prevent a sequential bottleneck. The scope of this paper is limited to dynamic load balancing only.

## IV. Dynamic Load balancing Techniques: A Brief Discussion

The circumstances determining the optimal balance change frequently or continuously during execution, so that the cost of the load balancing calculation after each change should be minimized in addition to optimizing the splitting of the actual calculation. This means that there must be a decision made every so often to decide if load balancing is necessary, and how much time to spend on it. Dynamic (or adaptive) policies, on the other hand, rely on recent system state information and determine the task assignments to processors at run time [8, 9, 10]. Hence, they are more attractive from a performance point of view [11, 12]. In the dynamic approach, the load balancing decisions are based on the current state of the system; tasks are allowed to move dynamically from an overloaded node to an under-loaded node to receive faster service. This ability to react to changes in the system is the main advantage of the dynamic approach to load balancing. A dynamic load balancing algorithm consists of four components, Load Measurement rule, an Information Exchange rule, an Initiation rule and a Load Balancing Operation.

## V. Policies in Load balancing Algorithms

Load balancing algorithms can be defined by their implementation of the following policies [13]:

i) **Information policy:** specifies what workload information to be collected, when it is to be collected and from where.
ii) **Triggering policy:** determines the appropriate period to start a load balancing operation.
iii) **Resource type policy:** classifies a resource as server or receiver of tasks according to its availability status.
iv) **Location policy:** uses the results of the resource type policy to find a suitable partner for a server or receiver.
v) **Selection policy:** defines the tasks that should be migrated from overloaded resources (source) to most idle resources.

## VI. Issues in Performance Evaluation

The main objective of load balancing methods is to speed up the execution of applications on resources whose workload varies at run time in unpredictable way [14]. Hence, it is significant to define metrics to measure the resource workload:

(i) How to measure resource workload?
(ii) What criteria are retaining to define this workload?
(iii) How to avoid the negative effects of resources dynamicity on the workload, and
(iv) How to take into account the resources heterogeneity in order to obtain an instantaneous average workload representative of the system?

## VII. Comparative Analysis of Various Dynamic Load Balancing Techniques

In this section we are going to present characteristic analysis of the different dynamic load balancing techniques based on the location of decision making, the information used for the decision making process, scalability factor, and the overhead of exchanging the profile information.

### i) Centralized Dynamic Load Balancing Techniques

In this technique, the responsibility of the Load balancing decision remains with the master node and the information used for the load balancing is gathered from the remaining (slave's) nodes on either on demand basis or after a certain predefined amount of fixed time interval, or even the information may be gather only when any change occurs in their working stage. The noticeable point is since the information is not send on arbitrarily, the unnecessary traffic over the interconnection network reduced. In addition, no unnecessary profile information is exchange overhead is encountered. But, the scalability remains limited with this technique.

### ii) Distributed Non-cooperative Dynamic Load Balancing Techniques

In distributed non-cooperative dynamic scheduling techniques the responsibility of the load balancing techniques distributed over all the working nodes i.e. workstations instead of the master node. The work load information is gathered based on the on demand basis i.e. whenever any node changes its current balanced working state to overloaded state, the specific node mat distribute the load information to reschedule to load to be balanced or alike. This technique provides moderate scalability over the centralized scheme. But since the load information has to distribute over several working nodes before rescheduling the current overload, this may increate the traffic in interconnection network in addition to the limited information exchange overhead.

### iii) Distributed Cooperative Optimal Dynamic Load Balancing Techniques

In distributed cooperative optimal dynamic load balancing techniques, unlike distributed non-cooperative dynamic scheduling techniques the responsibility of load balancing decision is scattered over all the workstations rather than master node. Further, in this case too load balancing information strategy is demand driven unlike the case of non-cooperative dynamic scheduling techniques with the exception of having average overhead during exchange of profile information. This technique does provide moderate scalability.

### iv) Distributed Cooperative Semi-optimal Heuristic Dynamic Load Balancing Techniques

Unlike the previous two techniques, in distributed cooperative semi-optimal heuristic dynamic load balancing techniques the responsibility load balancing decision is assigned over all the workstations collectively with demand driven information strategy and average profile information exchange overhead and moderate scalability.

### v) Distributed Cooperative Semi-optimal Approximation Dynamic Load Balancing Techniques

In this techniques too, the load balancing responsibility, information strategy and scalability remains same unlike in the case of distributed cooperative semi-optimal heuristic dynamic load balancing techniques i.e. to workstations, demand driven and moderate scalability respectively with the exception of consuming much more profile information exchange overhead increasing the traffic over the interconnection networks.

| Parameter | Dynamic Load Balancing Algorithms | | | | |
|---|---|---|---|---|---|
| | Centralized | Distributed | | | |
| | | Non-Cooperative | Cooperative | | |
| | | | Optimal | Semi-optimal | |
| | | | | Heuristic | Approximation |
| **Responsibility of Control** | Master Node | Each Workstation | All Workstations | All Workstations | All Workstations |
| **Information Strategy** | Demand Driven, Periodic, or State Change driven | Demand Driven | Demand Driven | Demand Driven | Demand Driven |
| **Scalability** | Limited | Moderate | Moderate | Moderate | Moderate |
| **Profile Information Exchange** | No | Limited | Average | Average | More |

Table 1. Comparative Analysis of Different Dynamic Load Balancing Techniques.

## VIII. Conclusion

From the above entire discussion we may conclude that, provided limited scalability is permitted, dynamic centralized load balancing techniques provides better performance over the alternatives as discussed being consuming no profile exchange overhead with master node assigning whole responsibility to take the load balancing decision with demand driven information strategy. But if system required sophisticated scalability, non-cooperative distributed dynamic load balancing techniques provides better solution as the overhead during profile information exchange is limited as compared to the other techniques within this group.

## IX. Future Work Plan

Near future, we are planning to enhance our research work by suggesting the ways to improve the performance of the dynamic load balancing techniques optimizing the limitations and constraints as discussed in the paper. We would like to further enhance our research work by determining the different alternative situations when to demand for load balancing information from the workstations to optimize the performance of the system.